\documentclass[12pt,a4paper]{article}
\usepackage{amsmath,amssymb}
\usepackage{graphicx}
\usepackage{rotating}
\usepackage{indentfirst}
\usepackage{setspace}
\usepackage{mathptmx}
\usepackage[left=4cm,right=2cm,top=2cm,bottom=2cm,bindingoffset=0cm]{geometry}
\usepackage{mathtext}
\usepackage{mathrsfs}
\usepackage[breaklinks=true,colorlinks=true,linkcolor=blue,urlcolor=blue,citecolor=blue]{hyperref}

\allowdisplaybreaks

\begin{document}

\title{{\bf Microscopic derivation of the Ginzburg-Landau equations for the periodic Anderson model in the coexistence phase of superconductivity and antiferromagnetism}}
\author{V.V. Val'kov, A.O. Zlotnikov\/\thanks{E-mail: zlotn@iph.krasn.ru}}
\date{}
\maketitle

\begin{center}
\textit{{\large Kirensky Institute of Physics, Russian Academy of Sciences, Siberian Branch, Krasnoyarsk, 660036}}
\end{center}

\doublespacing
\begin{abstract}
On the basis of the periodic Anderson model the microscopic Ginzburg-Landau equations for heavy-fermion superconductors in the coexistence phase of superconductivity and antiferromagnetism have been derived.
The obtained expressions are valid in the vicinity of quantum critical point of heavy-fermion superconductors when the onset temperatures of antiferromagnetism and superconductivity are sufficiently close to each other.
It is shown that the formation of antiferromagnetic ordering causes a decrease of the critical temperature of superconducting transition and order parameter in the phase of coexisting superconductivity and antiferromagnetism.

\end{abstract}

\par
\par

\clearpage

\section{Introduction}

The phenomenological Ginzburg-Landau (GL) theory is, up to now, widely used for describing new superconductors. Using the original Gor'kov method a microscopic derivation of GL equations of superconductors with anisotropic order parameter was done~\cite{Gorkov-63, Klemm-80, Riek-89}. Generalization of the Gor'kov method on the strongly correlated Hubbard model and Emery model in the atomic representation has been provided in~\cite{Zaitsev-88, Zaitsev-89}.  Microscopic GL equations have been also obtained for the t-J model in the slave-boson approach~\cite{Kuboki-98}.

In the last time a considerable interest is caused by a generalization of the GL theory for the materials in which at the same conditions the Cooper instability and long-range magnetic order are induced leading to a transition in the coexistence phase of superconductivity and antiferromagnetism (SC+AFM). For U-based ferromagnetic superconductors with triplet pairings the GL equations have been derived in~\cite{Mineev-04}. The microscopic GL expansion in the SC+AFM phase for the Hubbard model and t-J model in the mean-field approximation has also been given in Refs.~\cite{Kuboki-12, Kuboki-13}.

At present Ce-based intermetallic compounds with heavy fermions (for example, CeIn$_3$, CeRhIn$_5$, CePt$_2$In$_7$ \cite{Pfleiderer-09}) are widely known in which a transition to the homogeneous SC+AFM phase is implemented at low temperatures and under pressure. In this work in the framework of the effective periodic Anderson model describing in~\cite{VVV-12} the microscopic derivation of the GL equations for such superconductors  will be carried out.

\section{Model and Method}
The Hamiltonian of the periodic Anderson model, with the exchange interaction between localized electrons as the main mechanism of superconductivity and antiferromagnetism, has a form:
\begin{eqnarray}
\label{HamiltonianPAMeff}
\widehat{H}_{\text{eff}} & = & \sum_{m\sigma} \xi_{0} c_{m\sigma}^{\dag}c_{m\sigma} + \sum_{ml\sigma} t_{ml} c_{m\sigma}^{\dag}c_{l\sigma} +
\sum_{m\sigma} \xi_{\text{L} \sigma} X_{m}^{\sigma \sigma} + \frac{1}{2} \sum_{m \ne l} J_{ml} \left( \vec{S}_{m} \vec{S}_{l} -
\frac{1}{4}\hat{N}_{m}\hat{N}_{l} \right) +
\nonumber \\
& + & \sum_{ml\sigma} \left[ V_{ml}
c_{m\sigma}^{\dag}X_{l}^{0\sigma}
+ V^*_{ml}
X_{l}^{\sigma0}c_{m\sigma} \right],
\end{eqnarray}
where $c_{m\sigma}$ and $c_{m\sigma}^{\dag}$ are the annihilation and creation Fermi operators of an itinerant electron for the Wannier
site $m$ and the projection of the spin $\sigma$, $\xi_{0}$ is the site energy of an itinerant electron with respect to the chemical potential~$\mu$, $t_{ml}$ is hopping parameter between sites $m$ and $l$.
$X^{n s}_m$ is the Hubbard operator related to the site $m$, which is constructed using the atomic states $|m; n\rangle$ and $|m; s\rangle$ as usual, $\xi_{\text{L} \sigma}$ is the bare energy of a localized electron counted from $\mu$. $\vec{S}_m$ is the quasispin vector operator of a localized electron, whose components are related to the Hubbard operators by the standard way, $\hat{N}_{m}$ is the number operator of localized electrons, $J_{ml}$ is the effective exchange constant of quasilocalized electrons on sites~$m$~and~$l$.  $V_{ml}$ is the matrix element that describes the hybridization of the localized and itinerant
states related to the same Wannier site ($m = l$) or to different sites ($m \neq l$).

To describe antiferromagnetism a two-sublattice description is used. Therefore in the SC+AFM phase two normal and two anomalous Matsubara Green's functions~\cite{Zaitsev-88} for quasilocalized electrons are introduced:
\begin{eqnarray}
G^{(FF)}_{\sigma \sigma} \left( x,x'\right) = - \left \langle T_{\tau} \tilde{X}^{0 \sigma}\left(x\right) \tilde{X}^{\sigma 0} \left(x'\right)  \right \rangle,
G^{(GF)}_{\sigma \sigma} \left( x_1,x'\right) = - \left \langle T_{\tau} \tilde{Y}^{0 \sigma} \left(x_1\right) \tilde{X}^{\sigma 0} \left(x'\right)  \right \rangle,
\nonumber \\
\\
F^{\dag}_{FF \bar{\sigma}} \left( x,x'\right) = - \left \langle T_{\tau} \tilde{X}^{\bar{\sigma} 0} \left(x\right) \tilde{X}^{\sigma 0} \left(x'\right)  \right \rangle,
F^{\dag}_{GF\bar{\sigma}} \left( x_1,x'\right) = - \left \langle T_{\tau} \tilde{Y}^{\bar{\sigma} 0} \left(x_1\right) \tilde{X}^{\sigma 0} \left(x'\right)  \right \rangle.
\nonumber \\
\end{eqnarray}
Here $x = (\vec{R}_f,\tau)$, $x_1 = (\vec{R}_g,\tau)$, sites $f$ and $g$  belong to different sublattices, $\bar{\sigma}$ denotes opposite projection of spin $\sigma$. It should be noted that the energy of localized electron is renormalized by the antiferromagnetic exchange field. So the energy in the f-sublattice is  $\xi_{\text{L} \sigma} = \xi_{\text{L}} - \eta_{\sigma}J_0R_f/2$, where $\xi_{\text{L}} = E_0 - \mu - J_0n_{\text{L}}/4$, $n_{\text{L}} = \langle \hat{N}_{f} \rangle$, $R_f = \langle S_f^z \rangle$, $\eta_{\sigma} = +1 \, (-1)$ if $\sigma = \uparrow \, (\downarrow)$.

Following Gor'kov's method the equations of motion for the thermodynamic Green's functions have been simplified using the Hubbard-I approximation.  To derive the equations of motion on the mean-field level the Zwanzig-Mori projection-operator formalism has been used.

Using the relations between the Green's functions for itinerant electrons and localized Green's functions the Gor'kov equations for the Fourier transforms in the SC+AFM phase have been obtained:
\begin{eqnarray}
\label{G_Eq}
G^{(FF)}_{\sigma \sigma} \left( p \right) & = & G^{(FF)}_{A \sigma} \left( p \right) + \frac{\eta_{\sigma}}{F_{0 \sigma}^2} \Delta_{\vec{p}} G^{(FF)}_{A \sigma} \left( p \right) F^{\dag}_{GF\bar{\sigma}} \left( p \right) + \frac{\eta_{\sigma}}{F_{0 \bar{\sigma}}^2} \Delta_{\vec{p}} G^{(GF)}_{A \bar{\sigma}} \left( p \right) F^{\dag}_{FF\bar{\sigma}} \left( p \right),
\nonumber \\
\\
\label{F_Eq}
F^{\dag}_{GF\bar{\sigma}} \left( p \right) & = & \frac{\eta_{\bar{\sigma}}}{F_{0 \sigma}^2} \Delta_{\vec{p}} G^{(FF)}_{A \sigma} \left( -p \right) G^{(FF)}_{\sigma \sigma} \left( p \right)
+ \frac{\eta_{\bar{\sigma}}}{F_{0 \bar{\sigma}}^2} \Delta_{\vec{p}} G^{(GF)}_{A \bar{\sigma}} \left( -p \right) G^{(GF)}_{\sigma \sigma} \left( p \right).
\end{eqnarray}
The equations for the $G^{(GF)}_{\sigma \sigma} \left( p \right)$ and $F^{\dag}_{FF \bar{\sigma}} \left( p \right)$ can be obtained from the presented ones by substitution $G^{(FF)}_{A \sigma} \left( p \right) \to G^{(GF)}_{A \sigma} \left( p \right)$, and vice versa.  In equations~\eqref{G_Eq},~\eqref{F_Eq} it has been taken into account that $G^{(GF)}_{A \sigma}$ does not depend on spin.  Here $p = (\vec{p}, i\omega_n)$, $\omega_n$ is a Matsubara frequency for fermions, $F_{0 \sigma} = \left \langle X_f^{00} + X_f^{\sigma \sigma} \right \rangle$ is a Hubbard renormalization. Superconducting order parameter $\Delta_{\vec{p}}$  is given by integral equation:
\begin{equation}
\Delta_{\vec{p}} = \frac{T}{N}\sum_{k \sigma} \frac{J_{\vec{p}-\vec{k}}}{2} \eta_{\sigma} F_{GF \sigma}^{\dag} \left( k \right).
\end{equation}
In what follows the first-nearest-neighbor approximation for the exchange interaction is used. $G^{(FF)}_{A \sigma}$, $G^{(GF)}_{A \sigma}$ are related to Green's function of the background state: a N\'{e}el antiferromagnetic state if $R_f \ne 0$, a paramagnetic state if $R_f = 0$, and satisfied the following equations:
\begin{eqnarray}
\left( i\omega_n - \xi_{\text{L} \sigma} - F_{0\sigma}a_{p} \right) G^{(FF)}_{A \sigma} \left( p \right) & - & F_{0\sigma}b_p G^{(GF)}_{A \sigma} \left( p \right) = F_{0 \sigma},
\\
\left( i\omega_n - \xi_{\text{L} \bar{\sigma}} - F_{0\bar{\sigma}}a_{p} \right) G^{(GF)}_{A \sigma} \left( p \right) & - & F_{0\bar{\sigma}}b_p G^{(FF)}_{A \sigma} \left( p \right) = 0.
\end{eqnarray}
For simplicity we take into account only intra-sublattice hybridization~$V_{\vec{p}}$, then
\begin{equation}
a_p = \frac{\left(i\omega_n - \xi_{\text{c}\vec{p}}\right)|V_{\vec{p}}|^2}{\left(i\omega_n - \xi_{\text{c}\vec{p}}\right)^2-\Gamma_{\vec{p}}^2}, \, \, \, b_p = \frac{\Gamma_{\vec{p}}|V_{\vec{p}}|^2}{\left(i\omega_n - \xi_{\text{c}\vec{p}}\right)^2-\Gamma_{\vec{p}}^2},
\end{equation}
where $\xi_{\text{c} \vec{p}} = \xi_{0} + t_{\vec{p}}$, $t_{\vec{p}}$, $\Gamma_{\vec{p}}$ describe intra-sublattice and inter-sublattice hoppings in k-space.

\section{Results}
We limit ourself by the consideration of only d-wave superconductivity as leading pairing symmetry. Then in the absence of the external magnetic field GL equations
in the Wannier representation have a simple form:
\begin{eqnarray}
\label{GL_Eq_1}
\left[ \alpha_s \left( T \right) + \beta_s \Delta_f^2 + \gamma_{1} R_f^2 \right] \Delta_f = 0,
\\
\label{GL_Eq_2}
\left[ \alpha_m \left( T \right) + \beta_m R_f^2 + \gamma_{2} \Delta_f^2 \right] R_f = 0,
\end{eqnarray}
where $R_f$ is the antiferromagnetic order parameter as determined above, $\alpha_{s(m)} \left( T \right)$,
$\beta_{s(m)}$, $\gamma_{1(2)}$ --- GL coefficients, and the amplitude of superconducting order parameter at site $f$ is
\begin{equation}
\Delta_f = \sum_{\delta_x} \Delta_{f, f+\delta_x} - \sum_{\delta_y} \Delta_{f, f+\delta_y}.
\end{equation}
The summation on $\delta_x$, $\delta_y$ are carried over nearest sites along $x$ axis and $y$ axis, respectively. In real space GL equation can be obtained after taking a continuum limit for slowly varying $\Delta_f$, $R_f$.

The difference between the GL coefficients $\gamma_{1}$ and $\gamma_{2}$ is connected with the fact that Cooper instability doesn't change much the antiferromagnetic order parameter in the SC+AFM phase as it has been shown in~\cite{VVV-12}. Otherwise the antiferromagnetic ordering has a particularly strong influence on the superconducting order parameter. Therefore the term with~$\gamma_{2}$ in equation~(\ref{GL_Eq_2}) can be neglected.

It should be noted that equations~(\ref{GL_Eq_1}), (\ref{GL_Eq_2}) for non-zero $\Delta_f, R_f$ are valid only near a critical temperature $T_{\mbox{co}}$ of the SC+AFM phase. It means that the  N\'{e}el temperature $T_{\mbox{N}}$ should be close to the onset temperature of superconductivity $T_{\mbox{c}}$. We assume that $T_{\mbox{N}} > T_{\mbox{c}}$ which is a case of rare-earth heavy-fermion superconductors such as CeCu$_2$Si$_2$, CeIn$_3$, CeRhIn$_5$.

The temperature dependence of superconducting order parameter in this assumption is described by
\begin{equation}
\label{Delta_GL_AFM}
\Delta_f^2 \left( T \right) = - \frac{\alpha^{'}_s - \alpha^{'}_m \gamma_1/\beta_m}{\beta_s}\left( T - T_{\mbox{co}} \right),
\end{equation}
where $\alpha_s(T) = \alpha^{'}_s \left( T - T_{\mbox{c}} \right)$, $\alpha_m(T) = \alpha^{'}_m \left( T - T_{\mbox{N}} \right)$ and
\begin{equation}
T_{\mbox{co}} = \frac{\alpha^{'}_s - \alpha^{'}_m \gamma_1/\beta_m \cdot T_{\mbox{N}}/T_{\mbox{c}}}{\alpha^{'}_s - \alpha^{'}_m \gamma_1/\beta_m} T_{\mbox{c}}.
\end{equation}

When the antiferromagnetic order parameter is rather small near the critical temperature $T_{\mbox{co}}$, it is possible to expand the Green's functions $G^{(FF)}_{A \sigma}$, $G^{(GF)}_{A \sigma}$ into series in terms of $R_f$. Solving the system equations~(\ref{G_Eq}), (\ref{F_Eq}) by iterations in the SC+AFM phase microscopic expressions for the GL coefficients have been found:
\begin{equation}
\alpha_s \left(T\right) = 1 - \frac{T}{N}\frac{4J}{F_0^2} \sum_{k} \phi_{\text{d}}^2(\vec{k})\left[ G_0^{(FF)}(-k)G_0^{(FF)}(k) + G_0^{(GF)}(-k)G_0^{(GF)}(k) \right],
\end{equation}
where $F_0$ is a Hubbard renormalization and $G_0^{(FF)}$, $G_0^{(GF)}$ are Green's functions in the paramagnetic phase, $\phi_{\text{d}}$ --- d-wave basis function.
\begin{eqnarray}
\beta_s =  \frac{T}{N}\frac{16J}{F_0^6} \sum_{k} \phi_{\text{d}}^4(\vec{k})\left\{ \left[ G_0^{(FF)}(-k)G_0^{(FF)}(k) + G_0^{(GF)}(-k)G_0^{(GF)}(k) \right]^2 + \right.
\nonumber \\
\left. + \left[ G_0^{(FF)}(-k)G_0^{(GF)}(k) + G_0^{(GF)}(-k)G_0^{(FF)}(k) \right]^2 \right\}.
\end{eqnarray}
It is more convenient to represent the GL coefficients connected with antiferromagnetic ordering in the transformed form:
\begin{equation}
\alpha_m \left(T\right) = 1 - \frac{T}{N} \sum_{k} c\left(k\right), \, \, \, c\left(k\right) = \frac{\left(i\omega_n - \xi_{\text{L}} - 2F_0J\right)\left[ \left(i\omega_n-\xi_{\text{c} \vec{k}}\right)^2 - \Gamma_{\vec{k}}^2 \right]}{\prod\limits_{i}\left( i\omega_n-E_{i \vec{k}}^{(0)}\right)},
\end{equation}
\begin{eqnarray}
\beta_m & = & \frac{-T}{N} \sum_{k} \frac{4J^2 \left[ \left(i\omega_n-\xi_{\text{c} \vec{k}}\right)^2 - \Gamma_{\vec{k}}^2 \right] -
\left[ 4J\left(i\omega_n-\xi_{\text{c} \vec{k}}\right) - V_{\vec{k}}^2 \right] V_{\vec{k}}^2}{\prod\limits_{i}\left( i\omega_n-E_{i \vec{k}}^{(0)}\right)} c\left(k\right),
\end{eqnarray}
where $E_{i \vec{k}}^{(0)}$ --- hybridization spectrum in the paramagnetic phase in two-sublattice representation. And the coefficient arising in the SC+AFM state has a form
\begin{eqnarray}
\gamma_1 & = & \frac{-T}{N} \sum_{k} \frac{4J}{F_0^2} \phi_{\text{d}}^2(\vec{k}) \left\{ c(k)c(-k) - 2/F_0G_0^{(FF)}(-k)c(k) - 2/F_0G_0^{(FF)}(k)c(-k) + \right.
\nonumber \\
& + &  3/F_0^2G_0^{(FF)}(-k)G_0^{(FF)}(k) + 3/F_0^2G_0^{(GF)}(-k)G_0^{(GF)}(k) - G_0^{(FF)}(-k)d(k) -
\nonumber \\
& - & \left. G_0^{(FF)}(k)d(-k) - G_0^{(GF)}(-k)g(k) - G_0^{(GF)}(k)g(-k)  \right\},
\end{eqnarray}
\begin{eqnarray}
d(k) & = & \left( 2J -a_k + \frac{F_0b_k^2}{i\omega_n-\xi_{\text{L}}-F_0a_k} \right)\frac{G_0^{(FF)}(k)c(k)}{F_0},
\\
g(k) & = & \left[ \left( i\omega_n - \xi_{\text{L}} - F_0a_k  \right)^2 - F_0^2\left(  2J -a_k \right)^2 \right]\frac{\left(G_0^{(GF)}(k)\right)^2}{F_0^4b_k}.
\end{eqnarray}

\begin{figure}[htb!]\center
\includegraphics[height=6 cm]{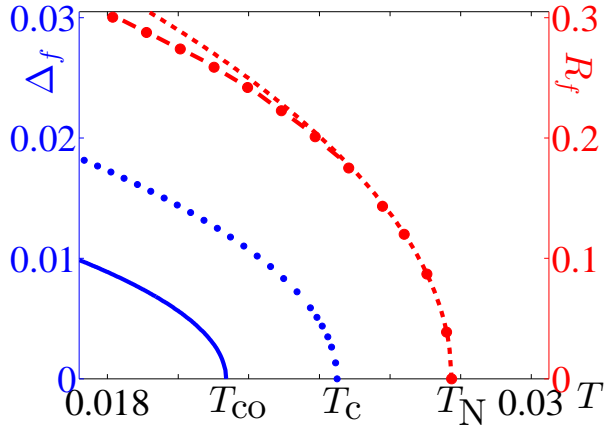}
\caption{Temperature dependence of the antiferromagnetic $R_f$ and superconducting $\Delta_f$ order parameters. The approximate dependence $R_f(T)$ from
Landau theory is shown by the dashed line, the exact dependence by the dash-dotted line, $T_{\mbox{N}}$ is the N\'{e}el temperature. The dotted line represents the Ginzburg-Landau dependence $\Delta_f(T)$ without antiferromagnetic ordering and $T_{\mbox{c}}$ is the onset temperature of superconductivity. The solid line is the same dependence in the coexistence phase of superconductivity and antiferromagnetism with the critical temperature $T_{\mbox{co}}$. \label{fig_R_D_T}}
\end{figure}

In Fig.~1 temperature dependencies of the antiferromagnetic~$R_f$ (dashed and dash-dotted lines) and superconducting~$\Delta_f$ (dotted and solid lines) order parameters obtained using the microscopic GL theory are plotted. We take parameters $n_{\mbox{e}} = 1.2$ (electron concentration), $E_0 = 1.8t_1$ (the bare energy of localized states), $V = 0.3|t_1|$, $J = 0.05|t_1|$, $t_1$ --- hopping matrix element of the tight-binding model. Comparing dashed and dash-dotted lines on the fig.~1 it is seen that the approximate dependence of the antiferromagnetic order parameter $R_f^2(T) = -\alpha_m(T)/\beta_m$ has a good agreement with the exact one obtained from the microscopic equations. The GL dependence $\Delta_f^2(T) = -\alpha_s(T)/\beta_s$ when antiferromagnetism is not taken into account is shown by the dotted line. The dependence $\Delta_f(T)$ in the SC+AFM phase is determined by the formula~(\ref{Delta_GL_AFM}) and shown on the Fig.~1 by the solid line. It is seen that antiferromagnetic ordering suppresses superconductivity and the onset temperature of superconductivity with the superconducting order parameter are decreased in an antiferromagnetic background.

\section{Conclusions}
We have derived the microscopic Ginzburg-Landau coefficients for strongly correlated heavy-fermion superconductors in the zero-field coexistence phase of superconductivity and antiferromagnetism.
The obtained coefficients have a strong dependence on the Fermi level position and the energy of localized level. They also are determined by the hybridization strength and magnitude of exchange interaction between localized electrons  which is responsible for both superconductivity and antiferromagnetism. The temperature dependencies of superconducting and antiferromagnetic order parameters near the transition temperature of the coexistence phase of superconductivity and antiferromagnetism have been found. It has been shown that superconductivity is suppressed due to antiferromagnetism.

\section{Acknowledgements}
This work was supported in part by the Russian Foundation for Basic Research, grant
no. 13-02-00523 and regional grant Siberia no. 15-42-04372. A.O. Zlotnikov acknowledges the support of RF
Presidential Grant no. SP-1370.2015.5.

\end{document}